# Graphene type dependence of carbon nanotubes/graphene nanoplatelets polyurethane hybrid nanocomposites: Micromechanical modeling and mechanical properties

Amir Navidfar[*], Levent Trabzon

*Faculty of Mechanical Engineering, Istanbul Technical University, Istanbul, Turkey;*
*MEMS Research Center, Istanbul Technical University, Istanbul, Turkey;*
*Nanotechnology Research and Application Center, Istanbul Technical University, Istanbul, Turkey;*

**Abstract**

Micromechanical modeling and mechanical properties of polyurethane (PU) hybrid nanocomposite foams with multi-walled carbon nanotubes (MWCNTs) and graphene nanoplatelets (GNPs) were investigated by mean of tensile strength, hardness, impact strength and modified Halpin–Tsai equation. Three types of graphene, with varied flake sizes and specific surface areas (SSA), were utilized to study the effect of graphene types on the synergistic effect of MWCNT/GNP hybrid nanofillers. The results indicate a remarkable synergetic effect between MWCNTs and GNP-1.5 (1:1) with a flake size of 1.5 μm and a higher SSA (750 $m^2$/g), which tensile strength of PU was improved by 43% as compared to 19% for PU/MWCNTs and 17% for PU/GNP-1.5 at 0.25 wt% nanofiller loadings. The synergy was successfully predicted using unit cell modeling, in which the calculated data agrees with the experimental results.

[*]**Correspondence to: Amir Navidfar (Email: Navidfar@itu.edu.tr)**

## 1.1 Introduction

Polyurethane (PU) foams, which are economic due to their low density, are frequently used in a widespread range of applications, such as insulation goals, automotive and electronic industries. However, their applications are limited because of their poor mechanical properties. Therefore, it seems attractive to modify PUs using nanoparticles to modify their mechanical properties [1-6]. In addition, components of PUs (polyol and isocyanate) are in a liquid form, which allow for the simple integration of solid nanofillers.

One-dimensional (1D) multi-walled carbon nanotubes (MWCNTs) and two-dimensional (2D) graphene nanoplatelets (GNPs) owing to their superior properties can be used as hybrid nanofillers to form well-dispersed three dimensional (3D) networks, which can overcome the dispersion problem of single nanofillers [7-12]. In order to solve the problem, acid functionalization of nanofillers can improve the dispersion, but the reaction conditions of acid oxidations are severe and may damage the graphitic structure [13]. Hybrid nanocomposites possess better mechanical properties in comparison with conventional nanocomposites that lead to the formation of an effective network for strain transferring [2, 14-16]. MWCNTs and graphene have a high ability to self-assemble due to $\pi-\pi$ interactions, which could inhibit aggregates resulting in enhancing the contact area between nanofillers and polymer matrixes [17, 18].

Recently, carbon nanotubes/graphene hybrid nanofillers were used in polymer nanocomposites to improve the mechanical properties [8, 16, 19, 20]. Weikang et al. [20] uniformly dispersed CNT-graphene hybrids into epoxy. Their results demonstrated that the tensile strength of the hybrid nanocomposite obtained an enhancement of 36%. Combining carbon nanotubes and graphene for improving the performance of epoxy nanocomposites was also studied by Shin-Yi Yang et al. [15]. The tensile strength and modulus of CNT+GNP/epoxy were increased by 14.5% and 22.6% at 1 wt% loading, respectively, which is obviously higher than the results of graphene/epoxy nanocomposites. The synergistic effect of the combinations did not completely understand. It is vital to determine the effect of nanofillers size on the properties of the hybrid nanocomposites. In addition, the ratio of the nanofillers is a significant parameter regarding the reinforcing capabilities of the nanocomposites [15, 21]. Chatterjee et al. [8] reported that the particle size of graphenes has a noticeable influence on the mechanical and thermal properties of the nanocomposites. They also exhibited that synergistic effects can be obtained using hybrid nanofillers especially for the CNT/graphene (9:1) and CNT/graphene (5:1), which are more effective than single nanofillers. Gao et al. [22] added two types of GNPs, xGnP-C750 (with an average diameter of 1 μm and a surface area of 750 $m^2/g$) and xGnP-M15 (a larger diameter of 15 μm but a lower surface area of 150 $m^2/g$) in PLA and indicated the highest reinforcement of 24% for 5 wt% xGnP-M15. However, in that study investigations were focused on single graphene nanofillers rather than on the effect of particle size on hybrid nanocomposites, which have not discussed in any detail.

A key question is which types of graphene are best suited to show a synergistic effect with MWCNTs to reinforcing polymeric nanocomposites. To investigate this, we compared three commercially available varieties of graphene with different flake sizes (24, 5 and 1.5 μm), aspect ratios and specific surface areas (150 and 750 $m^{2/}g$) to study synergistic effect of GNP with MWCNTs on tensile, hardness and impact properties of PU hybrid nanocomposites. We have observed a high synergistic effect and substantial improvement using GNP-1.5 with a lower flake size and a higher specific surface area at a low concentration of 0.25 wt%. Various nanofiller ratios were tested to find the optimal loading for mechanical properties and the highest synergy. The tensile strength of the single and hybrid nanocomposites was also compared with the predictions of the well-established Halpin-Tsai model, which was modified by adding an exponential shape factor, and the results fitted the experimental data successfully [23, 24]. In addition, a unit cell comprising graphene and MWCNTs was considered, which the synergistic effect was successfully predicted.

## 2. Materials and experimental setup

### 2.1 Materials

Fabrication of PUs was done by mixing the polyol and isocyanate in a weight ratio of 1:1.25, as recommended by the manufacturer according to Table 1. Graphene and MWCNTs were purchased from Nanografi Co.Ltd. MWCNTs were grown by chemical vapor deposition with an average diameter of 8 - 10 nm, the length of 1-3 µm, a specific surface area of 290 m$^2$/g and purity of more than 92%. GNP-24 refers to graphene nanoplatelets with a diameter of 24 µm, a thickness of 6 nm and a specific surface area of 150 m$^2$/g, according to the manufacturer datasheet. GNP-5 has a smaller diameter of 5 µm, a lower thickness of 3 nm and a specific surface area of 150 m$^2$/g. According to the manufacturer, GNP-1.5 has a smaller diameter of 1.5 µm, a thickness of 3 nm, but a higher specific surface area of 750 m$^2$/g.

Table 1. Properties of polyol and isocyanate components.

| Physical properties | Unit | Polyol | Isocyanate | Standards |
|---|---|---|---|---|
| **Density (25°C)** | g/cm3 | 1.11 | 1.23 | DIN 51 757 |
| **Viscosity (25°C)** | MPa.s | 600 ± 200 | 210 | ASTM D4878-98 |
| **OH content** | Mg KOH/g | 300 | - | ASTM D 4274-99 |
| **NCO content** | H$_2$O | - | %30.8 -%32 | ASTM 5155-01 |
| **Storage life** | Month | 3 | 6 | - |

### 2.2 Nanocomposites preparation

Prior to the synthesis of PU foams, 1.5 g of MWCNTs were added to 500 mL of 35% hydrogen peroxide (H$_2$O$_2$) at room temperature and mixed for 90 min. Subsequently, the solution was filtered and washed twice with distilled water to eliminate any H$_2$O$_2$ and then dried in an oven at 80°C for 12 h. Two sets of nanocomposites were fabricated with; (1) different contents of single MWCNTs and graphene, (2) different contents and ratios of carbon nanotubes/graphene hybrid nanocomposites at 0.25 to 0.75 wt% nanofiller contents. Nanofillers were added to the polyol and were stirred at 200-2000 rpm for 5 min. Then the mixture was ultrasonically dispersed for 5 min using an ultrasonic bath and stirred again at 2000 rpm for 5 min. Finally, the isocyanate was added to the nanofiller/polyol mixture and stirred for 20 s, as shown in Fig. 1. The details of the fabricated nanocomposites with MWCNTs, graphene, and their combinations are given in Table 2. The ratio of carbon nanotubes/graphene is vital for hybrid nanocomposites and thus specimens with three different ratios were fabricated for each graphene type.

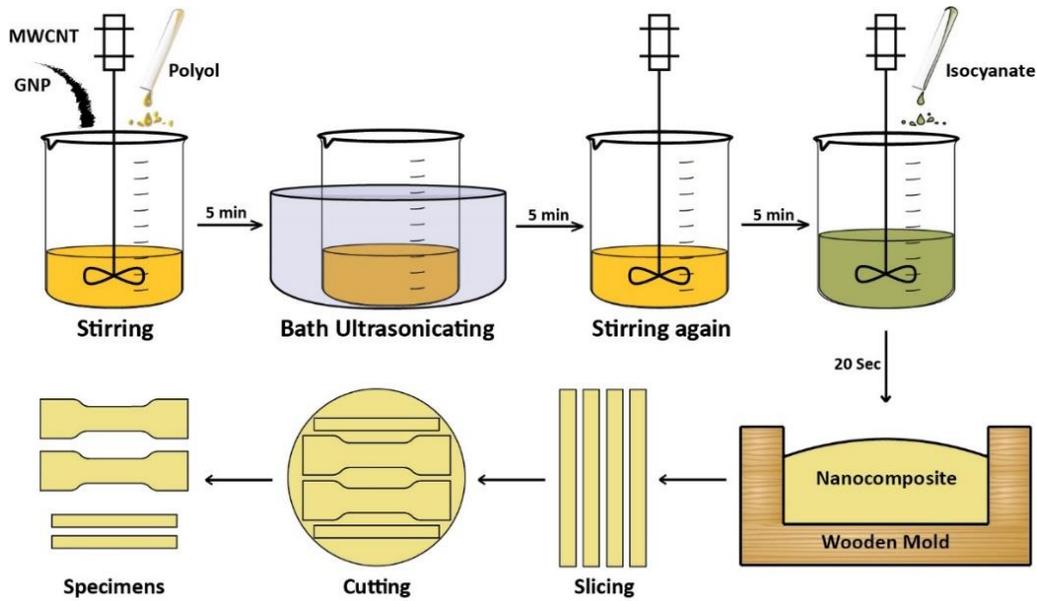

Fig. 1. The scheme of the nanocomposites fabrication steps.

Table 2. Levels of fabricated nanocomposites.

| Nanofillers | Concentrations (wt %) | | |
|---|---|---|---|
| **MWCNT** | 0.25 | 0.50 | 0.75 |
| **GNP-24** | 0.25 | 0.50 | 0.75 |
| **GNP-5** | 0.25 | 0.50 | 0.75 |
| **GNP-1.5** | 0.25 | 0.50 | 0.75 |
| **MWCNT+GNP-24** (1:3), (1:1), (3:1) | 0.25 | 0.50 | 0.75 |
| **MWCNT+GNP-5** (1:3), (1:1), (3:1) | 0.25 | - | - |
| **MWCNT+GNP-1.5** (1:3), (1:1), (3:1) | 0.25 | - | - |

### 2.3 Characterization and instruments

A turning machine was used for cutting slices of cured nanocomposites with a thickness of 10 mm. The slices were cut in a CNC machine according to ASTM 6110 for Charpy impact tests and ISO 1926 for tensile tests of rigid cellular plastics. Tensile properties were performed on at least five samples of each nanocomposite using Shimadzu, UTS machine equipped with a 1 kN load cell under a strain rate of 5 mm/min at room temperature. The impact strength of unnotched samples was obtained with a Devotrans Charpy impact machine. XF hardness tester is used to investigate Shore-0 hardness tests and at least six points of a sample were examined perpendicular to blowing direction. Raman spectroscopy was performed for the structure analysis of graphene nanoplatelets using a Renishaw inVia Raman spectrometer with a 532 nm laser. X-ray diffraction (XRD) analysis was performed on a RIGAKU Diffractometer using Cu (Ka) radiation. Thermogravimetric analysis was done under a nitrogen atmosphere on a Q600, TA Instruments with a heating rate of 5°C min$^{-1}$. Fourier transform infrared spectroscopy of the specimens was studied in the wavenumber range from 4000 to 650 cm$^{-1}$ at a resolution of 4 cm$^{-1}$ using a Thermo Scientific iS10 FTIR at room temperature. Dispersion states of hybrid nanofillers were investigated using Hitachi HighTech HT7700 transmission electron microscopy (TEM).

### 3. Results and discussion

### 3.1 Characterization

XRD spectra of graphene nanoplatelets are shown in Fig. 2(a). A typical (002) peak at $2\Theta = 26º$ is clear for all graphene while GNP-24 and GNP-5 have a much sharper peak than GNP-1.5, demonstrating that GNP-24 and GNP-5 are more crystalline than GNP-1.5. The structural defects of graphene play a critical role in the properties of nanofillers and their nanocomposites [25]. Raman spectra is a useful tool for the characterization of crystal structure, disorder and defects in graphene-based materials. The graphene nanoplatelets exhibit D-band and G-bands which indicates defects and the sp$^2$ carbon networks of the sample, respectively [22]. Compared with GNP-24 and GNP-5, Raman analysis (Fig. 2(b)) shows a high intensity of D-band at 1342 cm$^{-1}$ for GNP-1.5, suggesting more defects on the graphene sheets. Furthermore, the intensity ratio of D-band to G-band ($I_D/I_G = 0.49$) of GNP-1.5 is much higher than those of GNP-24 and GNP-5 ($I_D/I_G = 0.08$), representing more defects and porous graphene [26].

Fourier transform infrared spectra of neat PU and nanocomposites are illustrated in Fig 2(c). There are no visible changes in the FT-IR spectra of PU because MWCNTs and graphene display no obvious absorption in the infrared range [2]. Fig 2 (d) shows TGA results of PU and their nanocomposites with 0.25 wt% nanofiller loadings. It is apparent that graphene decelerate the thermal degradation and PU/GNP-1.5 indicates the highest thermal stability in comparison with other nanofillers due to the better dispersion in the PU matrix.

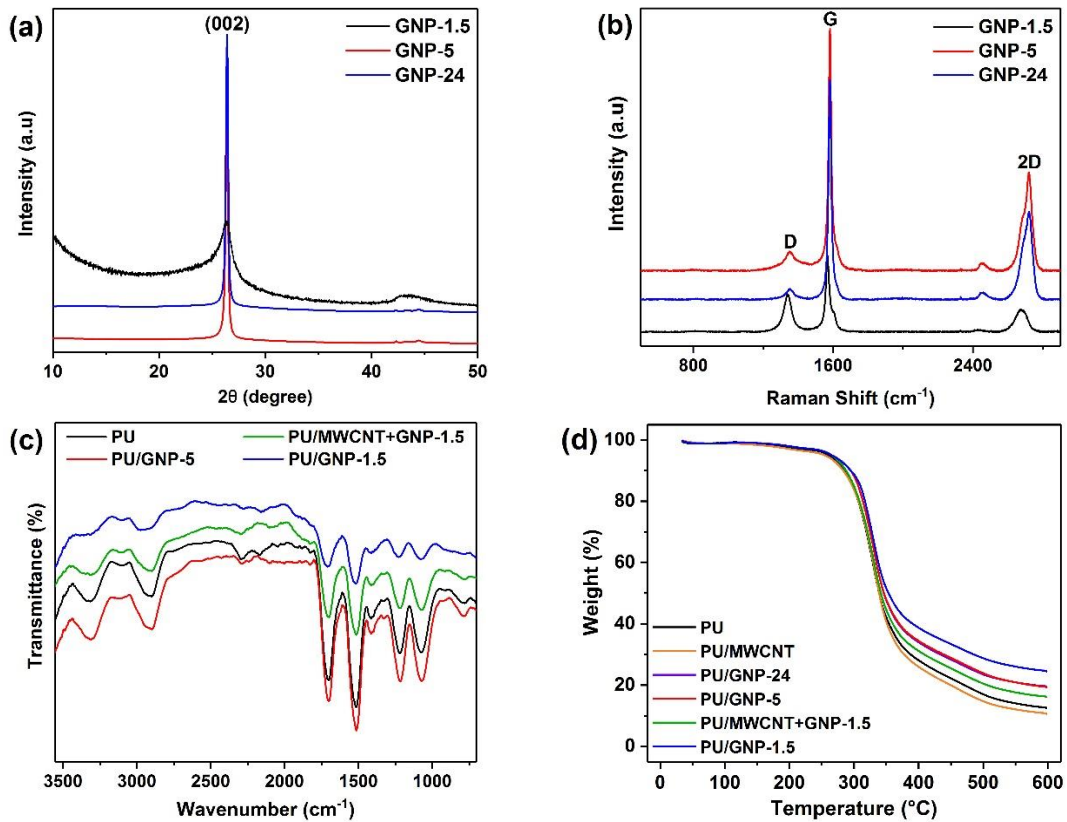

Fig. 2. (a) XRD patterns and (b) Raman spectra of graphene nanoplatelets, (c) FTIR spectra and (d) TGA curves of neat PU and nanocomposite foams.

## 3.2 Morphological characterization

The final properties of PU foams are severely dependent on its morphology, density, cell size, and walls thickness. In addition, the homogeneous dispersion of nanofillers in the polymer matrix is a key factor to fabricate reinforced polymer nanocomposites. Thus, it is very important to first characterize the microstructure of fabricated nanocomposites. Cross-sections of samples were fractured perpendicular to the foaming direction and fractured surfaces were coated with gold. Fig. 3 illustrates SEM images of neat polyurethane and PU nanocomposites with 0.25 wt% nanofiller content, which provide evidence of foams cellular microstructures with the average cell size values. It has been shown that the incorporation of the nanofillers decreased cell sizes of polyurethane, proving that even small amounts of nanofillers alter the foam's morphology. The decrease in cell size can also be attributed to the higher viscosity and nucleation effect of nanofillers, which have a profound influence on the mechanical properties of foams [27, 28]. As displayed in Fig. 3, the neat polyurethane has a uniform closed-cell structure while the cell structure is damaged through MWCNTs and graphene addition. The hybrid nanocomposite containing MWCNT+GNP-1.5 shows lowest cell size, which declares better dispersion [29]. Consequently, it is expected that the mechanical properties of nanocomposites with carbon nanotubes/graphene hybrids could be higher than others [2]. SEM micrographs provide visual evidence of the foam microstructure consisting of three phases of the polymer, nanofillers and bubbles. Fig. 3 also presents a series of microstructures focused on the strut area, in which carbon nanofillers are located. A number of agglomerations are visible in these areas while a better dispersion in hybrid nanocomposites with GNP-1.5 and MWCNTs (Fig 3 (f)) is obvious that 1D carbon nanotubes are well connected to 2D planar graphene floor via π-π stack interaction [30] and formed a 3D structure which inhibits aggregations. This 3D structure will enhance the contact surface areas between the polymer matrix and MWCNTs/graphene structures that is favorable to their mechanical properties.

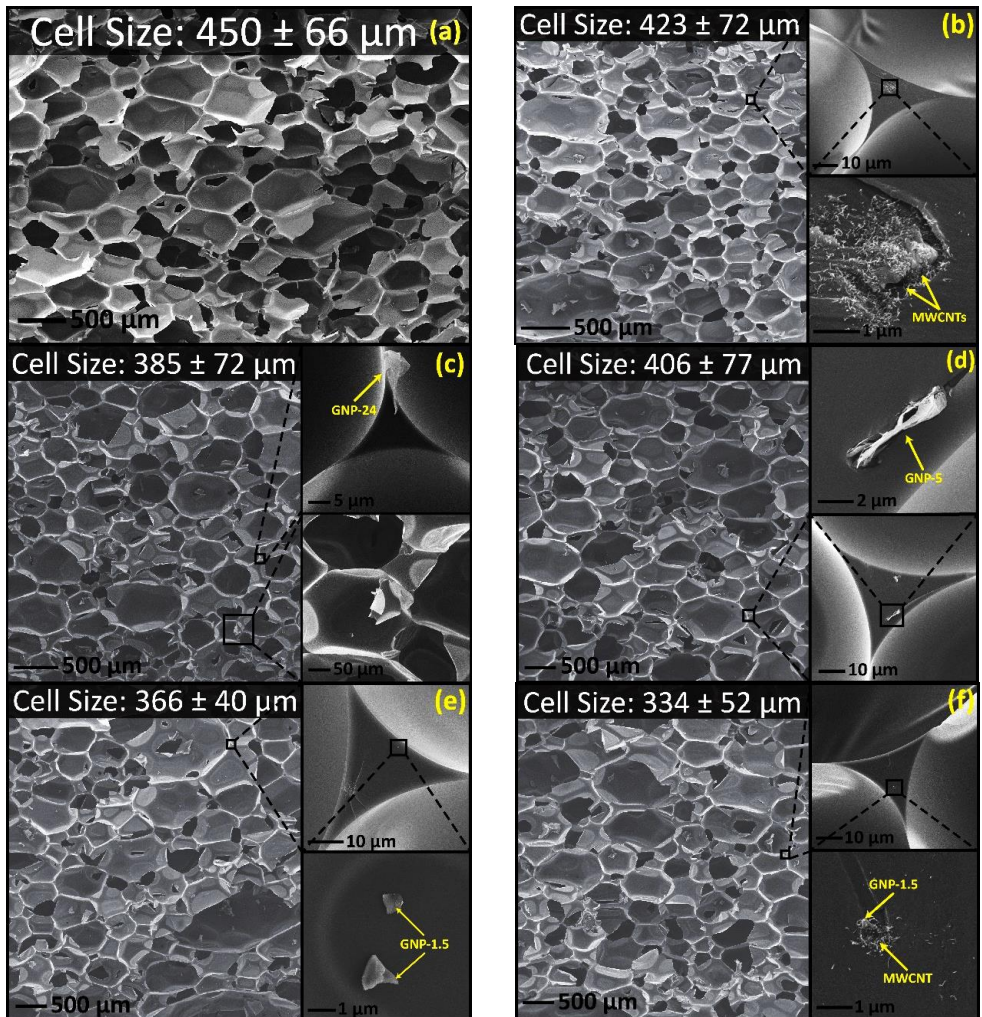

Fig. 3. SEM images of nanocomposite foams (0.25 wt%) (a) Polyurethane,
(b) PU/MWCNT, (c) PU/GNP-24, (d) PU/GNP-5, (e) PU/GNP-1.5,
(f) PU/MWCNT+GNP-1.5 (1:1).

## 3.3 Tensile Properties

Uniaxial tensile testing was used to examine the mechanical properties of neat polyurethane and its nanocomposites reinforced with carbon nanotubes and graphene nanoplatelets. Fig. 4(a) displays comparative results of MWCNTs and three types of graphene on ultimate tensile strength in various nanofiller loadings. The results show that carbon nanofillers are capable of improving the strength of PU in a low weight fraction. The tensile strength of neat polyurethane is about 0.39

MPa, whereas the addition of 0.25 wt% MWCNT, 0.25 wt% GNP-24 and 0.5 wt% GNP-1.5 enhanced the strength to about 0.468, 0.476 and 0.486 MPa with 19%, 21% and 24% improvement, respectively. The trend of strength improvement for MWCNTs is similar to GNP-24 and GNP-5, in which higher strength can be seen in the nanocomposites with 0.25 wt% nanofillers. The reinforcement efficiency of GNP-24 is obviously superior to that of GNP-5, which approves that a larger aspect ratio is beneficial for interfacial stress transfer from the matrix to graphene [31]. Results of a similar study by Valles et al. [32] showed that larger graphene provides better interfacial stress transfer with the polymer matrix, due to a more extensive contact area, which improves the mechanical properties. The nanocomposite containing 0.5 wt% GNP-1.5 has the highest tensile strength, which is attributed to the higher $I_D/I_G$ ratio and specific surface area of GNP-1.5 (750 m2/g) over those of GNP-24 and GNP-5 (150 $m^2/g$). The high specific surface area endows a better dispersion and an effective enhancement of mechanical properties in higher loadings [33]. The tensile strength of PU/MWCNTs is higher than PU/GNP-5 nanocomposites, while PU/GNP-24 shows better improvement. As a result, the superiority of graphene over MWCNTs depends on their properties such as aspect ratio, specific surface area. A study by Yan et al. [1] indicated more effective reinforcement of graphene than MWCNTs. In the contrary, Zakaria et al. [34] concluded that carbon nanotubes possess a better reinforcement effect. As can be seen in Fig. 4(a), MWCNTs, GNP-24 and GNP-5 reinforced PUs show a lower strength in higher loadings (0.5 and 0.75 wt%) because the dispersion becomes more challenging when nanofillers concentrations increased, which limits the improvement of mechanical properties [8]. Due to the better reinforcement effect of GNP-24 in comparison with GNP-5, GNP-24 was used for fabricating hybrid nanocomposites with MWCNT in different ratios to investigate synergistic effects of both nanofillers. The tensile strength of pure PU and PU/MWCNT+GNP-24 hybrid nanocomposites with fixed nanofiller contents (0.25, 0.5 and 0.75 wt%) are demonstrated in Fig. 4(b). As expected, a synergistic effect was observed for the strength where the nanocomposite with MWCNT+GNP-24 (1:3) showed the highest increase of 23% (0.483 MPa) relative to an increase of 12% and 15% for the single MWCNTs and GNP-24 based nanocomposites in 0.5 wt% loading, respectively. In nanocomposites with 0.75 wt% nanofiller loadings a similar trend was observed whereas the hybrid nanocomposite containing MWCNT+GNP-24 (1:3) showed the highest strength in this content. The higher tensile strength of hybrid nanocomposites clearly presents a synergistic effect. As presented by Yang et al. [15] carbon nanotubes could connect to graphene to form a 3D hybrid structure, which prevents aggregations of graphene nanoplatelets. 3D hybrid structure results in better interaction between hybrid MWCNT+GNP-24 and the polymer matrix, which a larger surface area and the increased contact area between hybrid nanofillers and the matrix could help to transfer the load in the tensile test [35, 36].

In this paper, three types of graphene nanoplatelets were used to investigate the effects of graphene size, specific surface area and their defects on the mechanical properties of hybrid nanocomposites. Fig. 4(c) represents a comparative result of graphene types on the ultimate tensile strength of hybrid nanocomposites at a constant level of 0.25 wt%. It is clear that the strength of MWCNT+GNP-1.5 hybrid nanocomposites is dramatically improved compared to the nanocomposites with other nanofillers. The tensile strength of nanocomposites with MWCNT+GNP-1.5 is increased up to about 43% (0.561 MPa) relative to that of PU. The fact that this is achieved at a nanofiller content of 0.25 wt% is remarkable. Whereas, there are moderate improvements in the PU/MWCNTs (~19%) and PU/GNP-1.5 (~17%) nanocomposites. Consequently, graphene nanoplatelets with a higher SSA and more defects have a greater ability to self-assemble with MWCNTs, which GNP-1.5 with MWCNTs exhibit a noticeable synergistic effect in reinforcing PU.

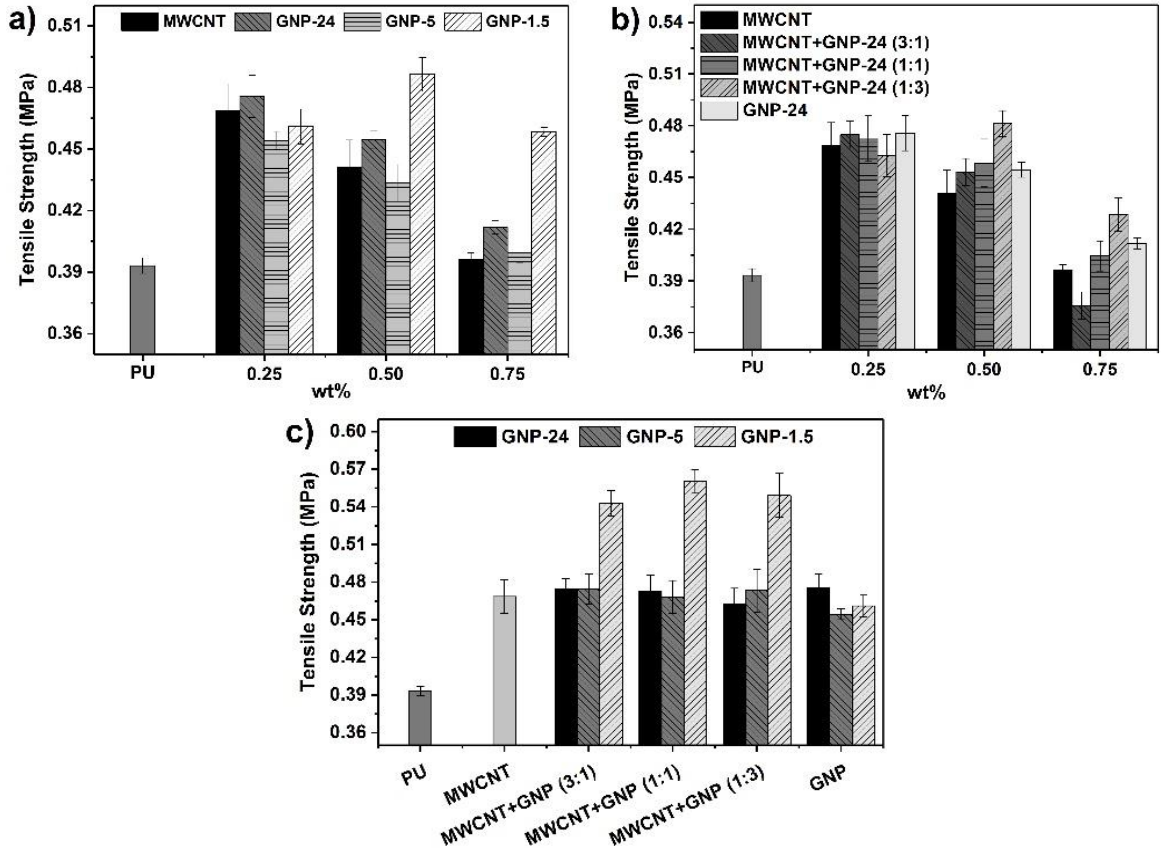

Fig. 4. The ultimate tensile strength of PU with (a) various nanofillers, (b) and (c) hybrid nanocomposites (0.25 wt%).

Lin Chen et al. [37] considered a unit cell with graphene/CNTs structure to calculate the effective thermal conductivity of the composites. A hybrid CNTs/graphene structure of the nanocomposites is displayed in Fig. 5 (a), which is similar to that of literature [37, 38]. TEM image of hybrid MWCNTs/GNP-1.5 (Fig. 5 (b)) approves the homogeneous distribution of MWCNTs on the GNP-1.5 surface to bridge the adjacent graphene, which formed a 3D interconnecting network. In our work, the unit cell is abstracted to predict the synergistic effect of graphene/CNTs hybrids, which is the periodic structure of the nanocomposites. Each unit cell comprises two half graphene with some MWCNTs between them and it has the same length and width of L and height of H. Both the GNPs and MWCNTs are considered as cylinders and are uniformly dispersed in the polymer matrix, in which the volume fraction of graphene in the unit cell equals to that in the nanocomposites, as follow:

$$V_{f,G} = \frac{V_G}{L^2 \cdot H} \tag{1}$$

Where, $V_{f,G}$ and $V_G$ are the volume fraction of graphene in nanocomposites and volume of single graphene, respectively. Assuming the graphene are located at the center of the cuboid as illustrated in Fig. 5, the following equating can be written [37]:

$$L = D_G + 2 \times \frac{1}{2} H_G = D_G + (H - t_G) \tag{2}$$

Where $D_G$ and $t_G$ are the diameter and thickness of graphene, respectively and $H_G$ corresponds to the distance between two half graphene. By simultaneously solving Eqs. (1) and (2), L, H and $H_G$ can be found. Since MWCNTs are distributed between graphene to form 3D hybrid structure, only carbon nanotubes with the length $L \geq H_G$ can attach to graphene in the unit cell. Used MWCNTs have an average length of 2000 nm. According to this model, nanocomposites with $H_G \leq 2000$ nm could show the synergistic effect in the tensile strength improvement of the nanocomposites with hybrid nanofillers, which outcomes of Table 3 and Fig. 4 prove precision of this model. As a result, the synergistic effect could be obtained by increasing the ratio and content of graphene in the nanocomposites, in which the distance between the graphene, $H_G$, decreases according to Eqs. (1) and (2). Thus, larger fractions of MWCNTs can connect graphene to form the 3D structure. Moreover, to achieve the synergistic effect in nanocomposites with larger flake graphene, either CNTs with a larger length should be used or the $H_G$ should decrease by enhancing the ratio and content of graphene.

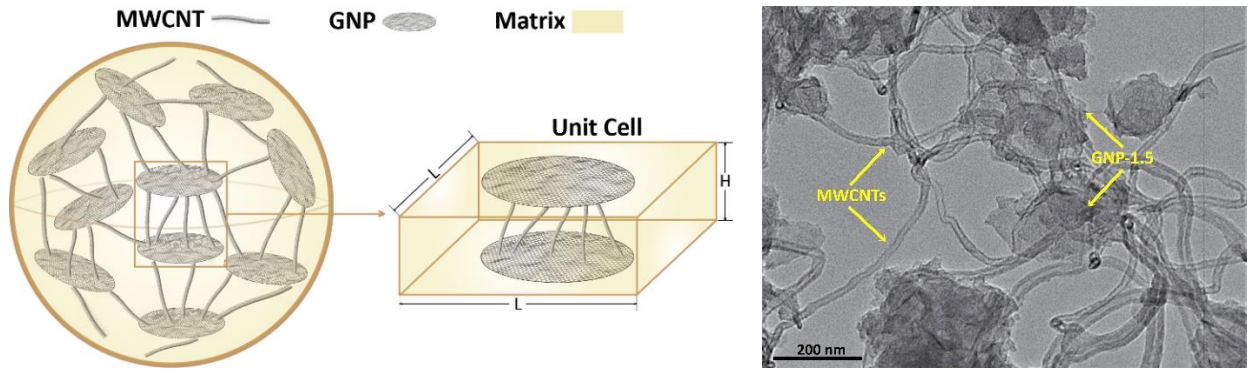

Fig. 5. a) Schematic diagram of the nanocomposite and a unit cell with a GNPs/MWCNTs structure, b) TEM images of GNP-1.5 and MWCNTs hybrids.

Table 3. Calculated $H_G$ of the nanocomposites, their synergistic effects and the strength enhancement.

| Loading | Nanofillers | Calculated $H_G$ (nm) | Synergy | Enhancement (%) |
|---|---|---|---|---|
| 0.25 wt% | MWCNT/GNP-24 (3:1) | 8061 | ✗ | 20.7 |
| | MWCNT/GNP-24 (1:1) | 4943 | ✗ | 20.2 |
| | MWCNT/GNP-24 (1:3) | 3618 | ✗ | 17.7 |
| | MWCNT/GNP-5 (3:1) | 4224 | ✗ | 20.6 |
| | MWCNT/GNP-5 (1:1) | 2887 | ✗ | 19 |
| | MWCNT/GNP-5 (1:3) | 2266 | ✗ | 20.3 |
| | MWCNT/GNP-1.5 (3:1) | **1640 (≤ 2000)*** | ✓ | **38** |
| | MWCNT/GNP-1.5 (1:1) | **1150 (≤ 2000)*** | ✓ | **43** |
| | MWCNT/GNP-1.5 (1:3) | **919 (≤ 2000)*** | ✓ | **40** |
| 0.50 wt% | MWCNT/GNP-24 (3:1) | 4939 | ✗ | 15.2 |
| | MWCNT/GNP-24 (1:1) | 2863 | ✗ | 16.6 |
| | MWCNT/GNP-24 (1:3) | **1992 (≤ 2000)*** | ✓ | **22.8** |
| 0.75 wt% | MWCNT/GNP-24 (3:1) | 3611 | ✗ | - 4.5 |
| | MWCNT/GNP-24 (1:1) | 2030 | ✗ | 2.9 |
| | MWCNT/GNP-24 (1:3) | **1388 (≤ 2000)*** | ✓ | **9** |

*Used MWCNTs have an average length of 2000 nm.

### 3.4 Impact Properties

Aside from tensile strength, impact properties are crucial in foam applications, which is related to fracture toughness [7]. Fig. 6(a) represents the impact strength of unnotched PU nanocomposite specimens. The maximum value of impact strength is achieved at 0.5 wt% loading for PU/MWCNTs nanocomposite with an enhancement of 21%, compared to that of pure polyurethane (0.794 kJ/m$^2$). Incorporating 0.25% weight fraction of GNP-24 and GNP-5 increased the impact strength of PU about 13.4% and 5.4%, respectively. Whereas, the maximum strength of GNP-1.5 reinforced nanocomposites can be seen at 0.75 wt% with an enhancement of 13.6%, which is attributed to the homogeneous dispersion of GNP-1.5 due to its higher specific surface area. Fig. 6(b) and (c) display the effects of hybrid nanofillers on the impact strength of PU nanocomposites. Synergistic effects cannot be seen in the impact strength, unlike tensile results. As reported in our previous works [2, 7, 39, 40], this is attributed to dominant reinforcement effect of MWCNTs on the impact strength due to their folding and shock absorbance properties [41].

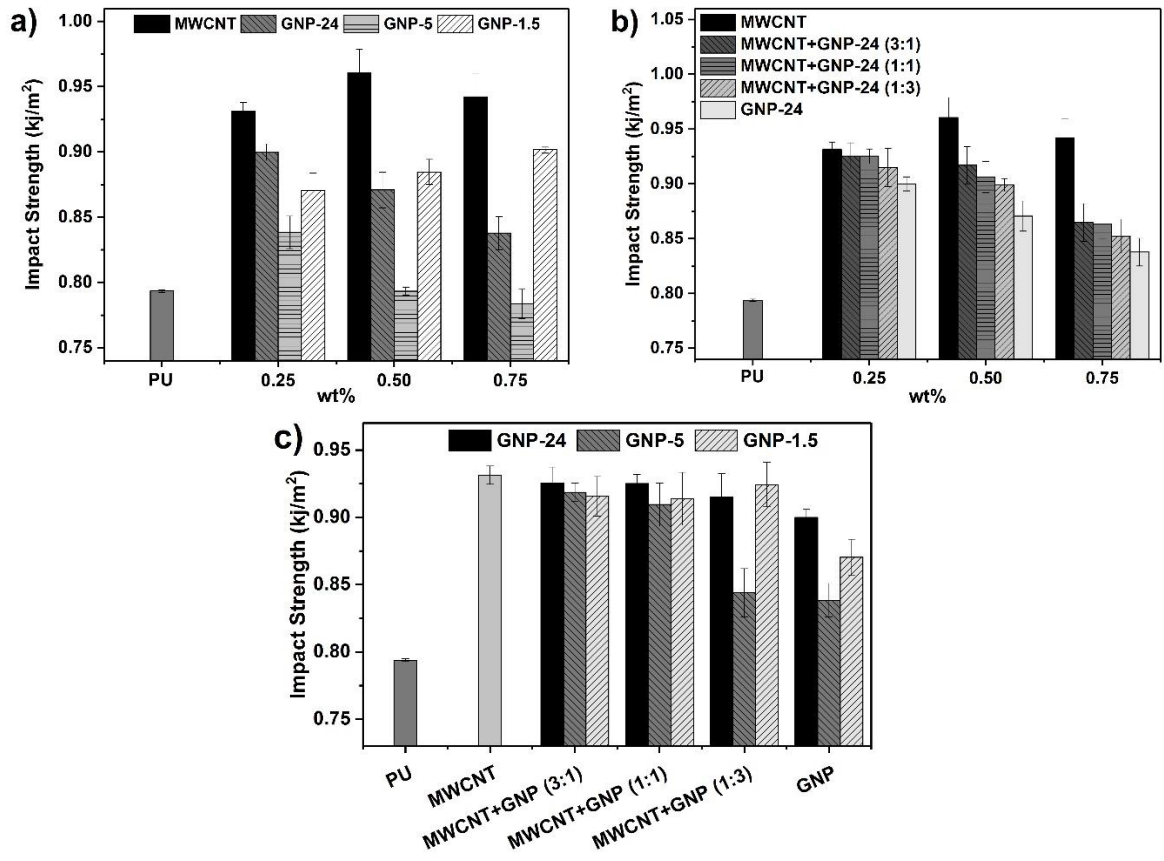

Fig. 6. The impact strength of PU with (a) various nanofillers, (b) and (c) hybrid nanocomposites (0.25 wt%).

### 3.5   Hardness

Fig. 7(a) illustrates the changes in Shore-0 hardness of PUs with graphene, MWCNTs and their different concentrations. The highest hardness values can be seen in the nanocomposites with 0.25 wt% GNP-24, MWCNT, GNP-1.5 and GNP-5 with 14%, 13.4%, 10.5% and 9.3% enhancements, respectively. This improvement in the hardness of nanocomposites may be due to the presence of a strong interaction between MWCNTs/graphene and PU. However, the hardness decreased in higher loading due to the uneven dispersion of nanoparticles. As mentioned in tensile results, agglomerations occurred in higher contents, which have an undesirable effect on the hardness. GNP-24 has a superior reinforcement effect in comparison with other nanofillers that approves a larger aspect ratio is advantageous for interfacial stress transfer from the matrix to nanofillers [31]. In the case of GNP-1.5, the higher specific surface area endows a better dispersion and an effective enhancement [33].

Nanocomposites show lower hardness in higher loadings (0.50 and 0.75 wt%) due to agglomerations. The hybrid nanocomposites were fabricated to overcome this challenge, which synergistic effect among nanofillers could reduce agglomerations. In Fig 7(b), it is observed that the value of hardness increases in hybrid nanocomposites. Polyurethane with MWCNT+GNP-24 (1:3) showed the maximum reinforcement of 15.4% compared to an increase of 10% and 11% with single MWCNTs and GNP-24 based nanocomposites in 0.5 wt% loadings, respectively. In nanocomposites with 0.75 wt% nanofillers, a similar trend was observed whereas the hybrid nanocomposite containing MWCNT+GNP-24 (1:3) indicates the highest hardness in this content. These results exhibited that MWCNTs and graphene showed the highlighted synergistic effect in enhancing the hardness of the nanocomposite foams.

Three types of graphene nanoplatelets were used to examine the influence of graphene size, specific surface area and defects on the mechanical properties of hybrid nanocomposites. Fig. 7(c) depicts comparative results of graphene types on the hardness of hybrid nanocomposites at a constant level of 0.25 wt%. It is observed that the hardness of MWCNT+GNP-1.5 hybrid nanocomposites is dramatically improved as compared to the nanocomposites with another graphene. The hardness of nanocomposites with hybrid MWCNT+GNP-1.5 is increased up to about 21% comparative to that of PU (41 Shore-0), while foams with single MWCNTs and GNP-1.5 improved the hardness about 13% and 10%, respectively. Therefore, graphene with a higher specific surface area and more defects facilitate synergistic effects and formation of 3D hybrid structures.

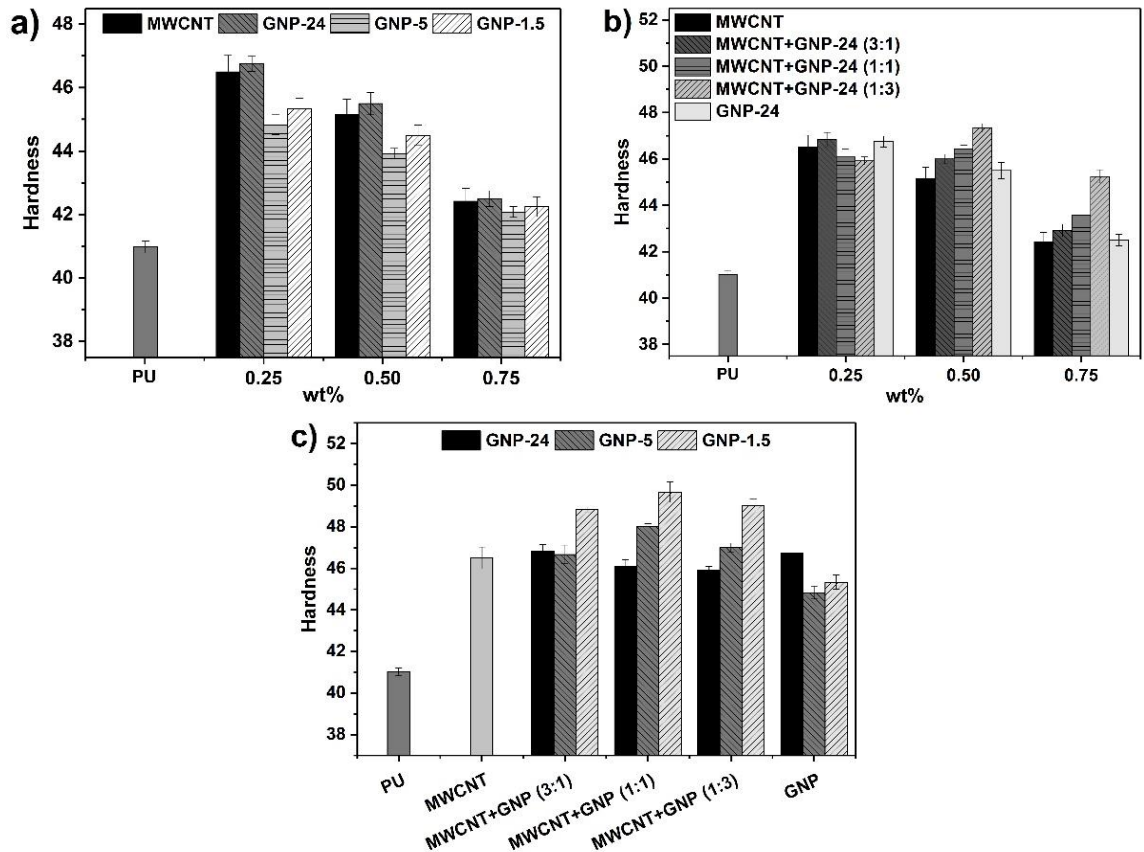

Fig. 7. Hardness of PU with (a) various nanofillers, (b) and (c) hybrid nanocomposites (0.25 wt%).

### 3.6 Micromechanical Analysis

The well-established Halpin-Tsai model, which can be widely utilized for predicting the tensile strength of randomly distributed nanofiller reinforced nanocomposites, was usually employed to calculate the theoretical tensile strength of the perfect dispersed nanocomposites. The experimental results of the nanocomposites are compared with the calculated values using the Halpin–Tsai equations as follows:

$$\sigma_C = \frac{3}{8}\sigma_L + \frac{5}{8}\sigma_T \tag{3}$$

Where $\sigma_C$, $\sigma_L$ and $\sigma_T$ are the nanocomposite, the longitudinal and transverse tensile strength, respectively. In our equations, graphene was assumed as effective rectangular fibers and MWCNTs were considered as discontinuous fibers, given by [42]:

$$\sigma_L = \left[\frac{1 + C\eta_L V_f}{1 - \eta_L V_f}\right]\sigma_{PU} \qquad \sigma_T = \left[\frac{1 + 2\eta_T V_f}{1 - \eta_T V_f}\right]\sigma_{PU} \tag{4}$$

In the above relations, $V_f$ is the volume fraction of MWCNTs and graphene, which can be calculated based on the density and weight fraction of nanofillers and neat PU [43]. Based on the manufacturer datasheet, the graphene and MWCNTs densities were considered 2250 kg/m³ and 2100 kg/m³, respectively. The density of PU matrix was measured 40.51 kg/m³. C is a constant shape factor related to the aspect ratio of fillers. For MWCNTs, $C_{CNT} = 2(l/d)$, where l and d refer to the average length and diameters of MWCNTs, which are about 2 μm and 9 nm, respectively. For graphene, $C_{GNP} = (2/3)(l/t)$, where l and t refer to the diameter and thickness of graphene. The parameters $\eta_L$, $\zeta_L$, $\eta_T$ and $\zeta_T$ are given by;

$$\zeta_L, \eta_L = \frac{(\sigma_f/\sigma_{PU}) - 1}{(\sigma_f/\sigma_{PU}) + C} \qquad \zeta_T, \eta_T = \frac{(\sigma_f/\sigma_{PU}) - 1}{(\sigma_f/\sigma_{PU}) + 2} \tag{5}$$

Where $\sigma_f$ and $\sigma_{PU}$ are tensile strengths of nanofillers and the polyurethane matrix, respectively. The parameters $\eta_L$ and $\eta_T$ were used for MWCNTs, while $\zeta_L$ and $\zeta_T$ were utilized for graphene nanocomposites. The tensile strength of PU is $\sigma_{PU} = 0.3933$ MPa and the strength of MWCNTs and graphene are assumed 63 GPa and 130 GPa, respectively [44]. The constant shape factor C was modified as an exponential shape factor ψ to fit nonlinear region for higher nanofiller contents due to agglomerations of nanofillers, which Halpin-Tsai equation was not assumed. The exponential shape factor ψ has the formula [23]:

$$\psi = CK = Ce^{-aV_f - b} \tag{6}$$

The constants a and b are the degree of nanofiller aggregations, accounting for the nonlinear behavior of the Halpin–Tsai equation and are found with respect to experimental results. The modified Halpin-Tsai equation may be rewritten as follows:

$$\sigma_L = \left[\frac{1 + \psi\eta_L V_f}{1 - \eta_L V_f}\right]\sigma_{PU} \qquad \zeta_L, \eta_L = \frac{(\sigma_f/\sigma_{PU}) - 1}{(\sigma_f/\sigma_{PU}) + \psi} \tag{7}$$

After an organized variation of aggregation-related a and b constants, ψ relations become equal to:

$$\psi = C_{CNT} e^{-1030.56 V_f + 1.26} \qquad \text{For PU/MWCNT} \tag{8a}$$
$$\psi = C_{GNP} e^{-905.64 V_f - 0.63} \qquad \text{For PU/GNP-24} \tag{8b}$$
$$\psi = C_{GNP} e^{-1048.93 V_f + 0.817} \qquad \text{For PU/GNP-5} \tag{8c}$$
$$\psi = C_{GNP} e^{-444.43 V_f + 0.696} \qquad \text{For PU/GNP-1.5} \tag{8d}$$

Experimental results and theoretical fits by Halpin-Tsai equation (Eqs. (4) and (5)) and modified Halpin-Tsai equation with the exponential shape factor (Eqs. (7) and (8)) were plotted in Fig. 8. Comparisons indicate that there is an obvious difference between the experimental strength results and Halpin–Tsai equation, while the predictions of the modified Halpin–Tsai equation correlates well with the experimental data. It is noteworthy that the experimental results of nanocomposites with GNP-24 and GNP-5 are far from the Halpin-Tsai equation (blue lines in Fig. 8) in comparison with PU/GNP-1.5, due to the higher folding and breaking tendency of larger graphene. Gao et al. [22] reported that graphene with a larger flake is more susceptible to shortening during mixing and to bending after combining, which could lower their aspect ratio even more. The modified Halpin–Tsai with exponential shape factor is a semi-empirical method that can be utilized to calculate the results slightly outside of tested nanofillers concentration considered by extrapolation.

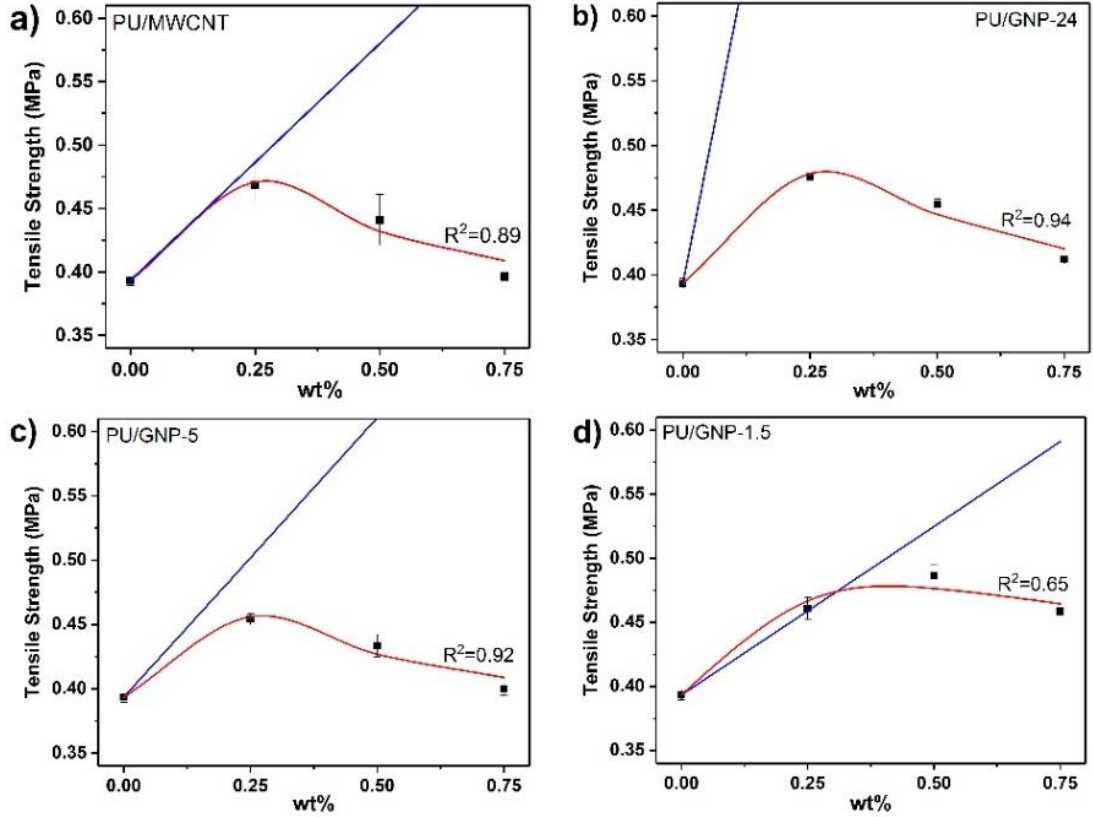

Fig. 8. Experimental data (■), Halpin-Tsai (blue) and modified Halpin-Tsai equations (red), for (a) PU/MWCNT, (b) PU/GNP-24, (c) PU/GNP-5, (d) PU/GNP-1.5 nanocomposites.

For hybrid nanofillers reinforced nanocomposites Halpin–Tsai equation could be rewritten as [16]:

$$\sigma_L = \left[\frac{1 + C_{CNT}\eta_L V_{CNT} + C_{GNP}\zeta_L V_{GNP}}{1 - \eta_L V_{CNT} - \zeta_L V_{GNP}}\right]\sigma_{PU}, \quad \sigma_T = \left[\frac{1 + 2\eta_L V_{CNT} + 2\zeta_L V_{GNP}}{1 - \eta_L V_{CNT} - \zeta_L V_{GNP}}\right]\sigma_{PU} \quad (9)$$

In addition, the modified Halpin-Tsai equation can be used for hybrid nanocomposites using ψ instead of $C_{CNT}$ and $C_{GNP}$ in Eq. 9. For example, ψ for the modified Halpin-Tsai model for the hybrid nanocomposite with MWCNT+GNP-24 (1:1) equals to:

$$\psi = Ce^{-846.36V_f - 0.141} \quad (10)$$

Fig. 9(a) shows experimental results, Halpin-Tsai equation and modified Halpin-Tsai equation with the exponential shape factor (Eq.10). The predictions of the modified Halpin–Tsai equation correlates well with the experimental results.

As mentioned in Fig. 4(c), MWCNT+GNP-1.5 hybrid nanofillers displayed an outstanding synergistic effect on tensile strength, which can be mostly attributed to the homogeneous 3D dispersion of MWCNTs and GNP-1.5 in the PU matrix. 3D distribution of MWCNT+GNP-1.5 hybrid nanofiller can be defined by the Halpin-Tsai equation. Fig. 9(b) reveals that the Halpin-Tsai equation under-predict the experimental data of the nanocomposites with MWCNT+GNP-1.5 hybrid nanofillers, while the experimental values of the single PU/MWCNTs, PU/graphene and other hybrid nanofillers are generally lower than their values estimated using the Halpin-Tsai model. As shown in Fig. 9(b), the experimental results of MWCNT+GNP-1.5 hybrid nanofillers are closer to the longitudinal tensile strength of the theory (black columns), which approves the 3D uniform dispersion [16]. In this case, a higher reinforcement effect is achieved by adding MWCNT+GNP-1.5 to PU matrix, which provides a synergistic effect and a higher tensile strength [45-47].

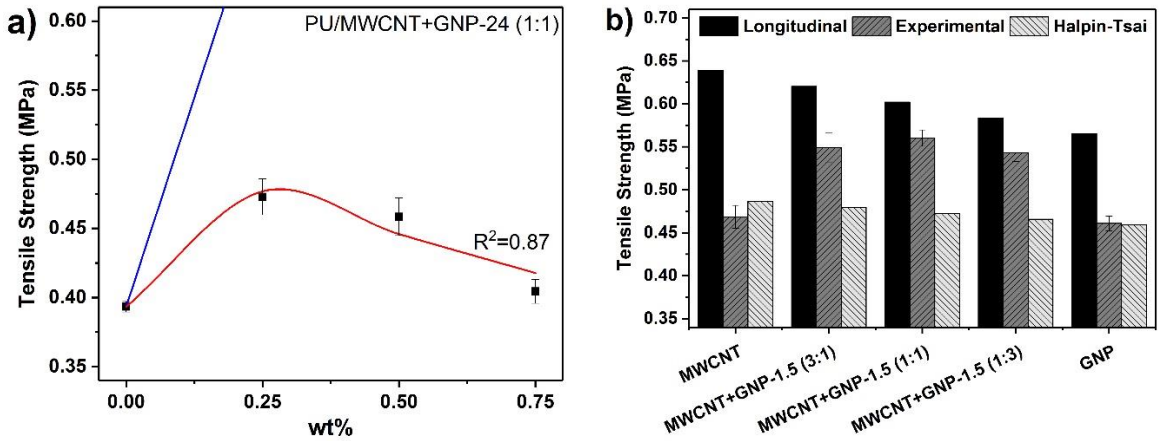

Fig. 9. (a) Experimental data (■), Halpin-Tsai (blue) and modified Halpin-Tsai equations (red) for PU/MWCNT-GNP-24 (1:1) and (b) Longitudinal, Halpin-Tsai model and experimental data for PU/MWCNT-GNP-1.5 hybrids

## 4. Conclusion

In summary, graphene type dependence of three-dimensional graphene/MWCNTs hybrid nanofillers on mechanical properties improvement of polyurethane nanocomposites was studied using tensile test, impact strength, hardness and micromechanical modeling. Three types of graphene with various flake sizes and specific surface areas were used to achieve a synergistic effect at 0.25 to 0.75 wt% loadings. MWCNTs and GNP-1.5 (1:1) with a flake size of 1.5 μm and a higher SSA (750 $m^2$/g) exhibited the highest synergistic effect at a low nanofiller content of 0.25 wt%, which the tensile strength was improved about 43%, in comparison with pure PU. Modified Halpin-Tsai modeling with an exponential shape factor was used to fit with the tensile strength results of the single and hybrid nanocomposites. A unit cell containing graphene and MWCNTs structure was introduced to predict the optimal ratio and content of the nanofillers in hybrid nanocomposites to attaining the synergism, which the modeling results were extremely compatible with the experimental data. The connections between the graphene and MWCNTs were more possibly formed when the length of carbon nanotubes is higher than the gap between graphene in the unit cell. The presented model could provide guidance for achieving the synergistic effect between different CNTs and graphene without many experiments.